\documentstyle[a4,psfig,epsf]{article}

\newcommand{\beq}{\begin{equation}}
\newcommand{\eeq}{\end{equation}}
\begin{document}

% @twocolumnfalse\endcsname

\title{Shock propagation in a granular chain}
\author{Erwan Hasco\"et$^{1}$, Hans J. Herrmann$^{1,2}$
and Vittorio Loreto$^1$}
\date{\today}
\pagestyle{plain}
\maketitle
\centerline{
$^1$P.M.M.H. Ecole Sup\'erieure de Physique et Chimie Industrielles,}
\centerline{10, rue Vauquelin, 75231 Paris CEDEX 05 France}
\centerline{$^2$ICA1, University of Stuttgart, Pfaffenwaldring 27, D-70569
Stuttgart Germany}
\begin{abstract}
We numerically solve the propagation of a shock wave in a chain of
elastic beads with no restoring forces under traction (no-tension
elasticity). We find a sequence of peaks of decreasing amplitude and
velocity. Analyzing the main peak at different times we confirm a
recently proposed scaling law for its decay. 
\end{abstract}
\smallskip
{\small Key words: Granular Media, Nonlinear waves}

\newpage
\section{Introduction}

The study of granular matter is of interest for fundamental physics as
well as for technological applications. The intensive studies of
liquids and solids gave physicists powerful tools for investigating
these states of matter. These tools are however very difficult
to apply to granular matter \cite{JNB}.

Few works have been done on sound propagation in real granular
media. Experiments have been performed by Liu and Nagel
\cite{Liu,LiuNagel1,LiuNagel2} who investigated the propagation of low
amplitude vibrations in a box of spherical beads. It was concluded
that non-linearity and disorder make wave propagation behave in very
unexpected ways: the fragile structure of contacts between grains make
them sensitive to rearrangements changing dramatically the amplitude
response of the receptor to the source.  Besides, disorder gives rise
to important interference effects that can lead to localization
\cite{Sornette}. Numerical simulations have been made by Melin
\cite{Melin} who studied the depth dependence of sound speed in
granular media. The results obtained were different from the power law
behavior between sound speed and pressure predicted by Goddard
\cite{Goddard} in an effective medium calculation.

Three dimensional models seem to be very difficult to investigate
 directly and we need to begin with simplified models in order to
 isolate specific features of real granular media. In this paper we
 try to understand the effect of non-linearity on wave propagation in
 one dimension. A previous study was done by Nesterenko
 \cite{Nesterenko} on a chain of spherical beads obeying the elastic
 Hertz law of contact. It was shown both analytically and numerically
 that the chain submitted to compression pulses involves solitary wave
 propagation. This has been confirmed by the experiments of Lazaridi
 and Nesterenko \cite{LazaridiNesterenko} and of Coste, Falcon and
 Fauve \cite{CFF}.

Here, we want to study the case of a chain of beads initially in
contact and submitted to a shock perturbation. The beads interact via
an elastic contact law only when they are compressed. The problem is
therefore highly non-linear because as soon as there are broken
contacts the chain behaves as an ensemble of independent elastic
systems.  We can look at this problem even in terms of a linear chain
of points connected by springs with completely asymmetric elastic
constants: $k$ in compression and zero in extension. Here we must
precise that the Hertz law which has been studied until now correspond
to perfect spherical beads. In a real granular medium the shape of the
beads at contacts can be far from spherical. In fact beads
interactions can be modelized with a force law exponent varying from
one to two. Our linear compression force model belongs to this
range. This one-dimensional system is clearly far from being realistic
but it exhibits some features which can help understanding of the more
general problem. In particular we try to elucidate how the front wave
propagates and how its lost energy can contribute to the formation of
several other secondary waves.  The outline of the paper is as
follows: in section 2 we define our model and we present the setup for
the numerical simulations.  Section 3 is devoted to the description of
the results concerning the phenomenology of the system. In section 4
we discuss in detail the mechanism of front formation.  Discussions
and conclusion will be given in section 5.

\section{Model definition}

\begin{figure}[htbp]
\centerline{
        \psfig{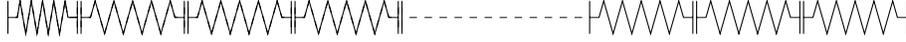}}
\caption[]{Simple experimental realization of the model.}
\label{fig}
\end{figure}

Our model is composed of $N$ spherical beads of radius $R$ and mass
$m$. We define the force between two neighboring beads as varying
linearly under compression and being equal to zero when the beads are
not in contact. It can be written as: \beq F=k \delta\theta(\delta)
\eeq where $k$ is the spring constant, $2R-\delta$ the distance
between the centers of the two neighboring beads and $\theta$ the
Heaviside function. Mechanically such systems are known as
``no-tension elastic'' \cite{AA}.  Initially all the beads are just
touching each other without exerting any forces on each other except
for the first bead which penetrates the second one by $\delta_0$ . We
then let the system evolve with the left end fixed and the right end
free. An experimental realization of the model can easily be done as
it can be seen from the Fig.(\ref{fig}). 
This one dimensional array of non-connected linear springs is the
simplest model for no-tension elasticity.

\noindent The equations of motion are the following:
\begin{equation}
     m\ddot{u}_i=k\delta_{i-1,i}\theta(\delta_{i-1,i})-
k\delta_{i,i+1}\theta(\delta_{i,i+1})\quad 0 < i < N-1
\end{equation}
where $u_i$ is the displacement of bead number $i$.
\noindent One can write:
\begin{eqnarray}
   \delta_{i-1,i} & = & 2R-(x_i-x_{i-1}) \\
                 &  = & 2R-(u_i+x_{i,0}-(u_{i-1}+x_{i-1,0})) \\
                 &  = & u_{i-1}-u_i 
\end{eqnarray}
where $x_{i}$ is the position of bead number $i$
and $x_{i,0}=(2i+1)R$.

\noindent Thus,
\begin{equation}
     m\ddot{u}_i=k(u_{i-1}-u_i)\theta(u_{i-1}-u_i)-k(u_i-u_{i+1})
\theta(u_i-u_{i+1})\quad 0 < i < N-1
\end{equation}
with the boundary conditions given by
\beq
\left\{
\begin{array}{l}
u_0=\delta_0 \,\, \mbox{for} \,\,t\geq 0 \\
u_i=0 \,\,\mbox{for} \,\,i>0 \,\,  \mbox{and} \,\,t=0
\end{array}
\right.
\eeq
These boundary conditions correspond to the study of the propagation
of a shock in the chain.
From now onwards we will denote with forward and backward direction
the direction of increasing and decreasing $i$, respectively.
For the numerical implementation of the analysis we have chosen 
$N=1500$, $R=5mm$,
$\rho=1.9\times10^3 kg.m^{-3}$, $k=10^7 N.m^{-1}$ and
$\delta_0=0.5mm$.  Several algorithms have been used to solve the
system of equations (6). First, we implemented the Verlet scheme which
was not able to give a good enough precision. Much better accuracy was
obtained with a 5th order Gear predictor-corrector method being even
more precise than the 4th order Runge-Kutta scheme. The test of the
accuracy was obtained by monitoring the energy conservation
of the system. During the computation, energy conservation was 
satisfied with a relative error lower than $10^{-6}$ with a time step 
$\Delta t=10^{-8}s$. It is worth to stress that during the total
time of evolution the perturbation never reaches the end of the chain.

\section{Front phenomenology and comparison with the harmonic case}

\begin{figure}[htbp]
\centerline{
       \psfig{figure=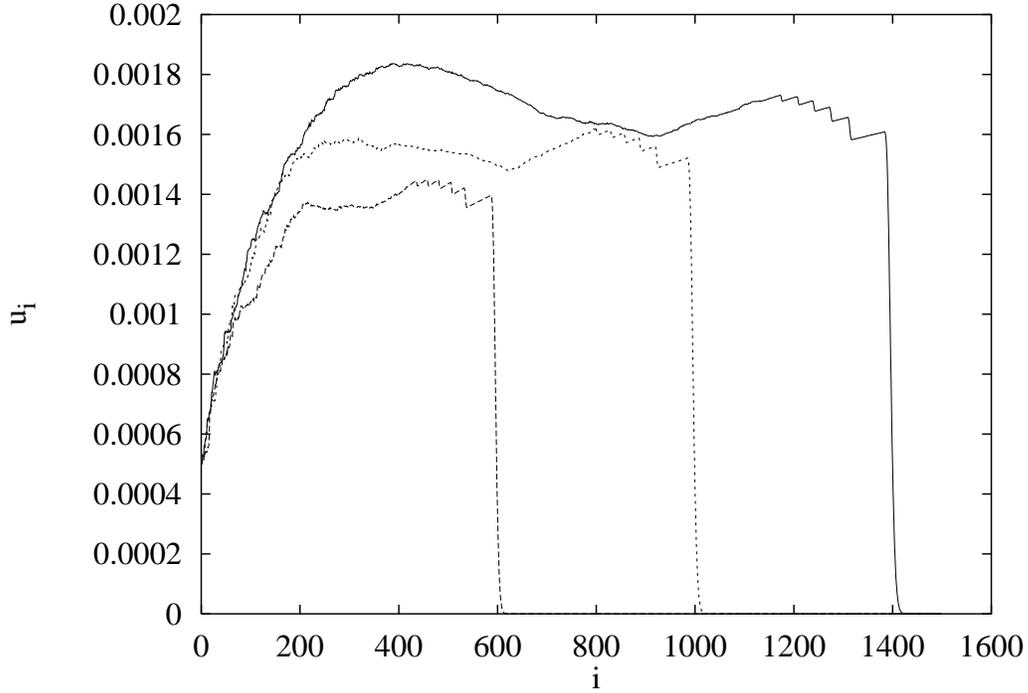,width=14cm}
}
\caption[]{Displacements of the beads at propagation times
  equal to $6\times10^{-3}s$, $10^{-2}s$ and $1.4\times10^{-2}s$.}
\label{fig1}
\end{figure}

\begin{figure}[htbp]
\centerline{
        \psfig{figure=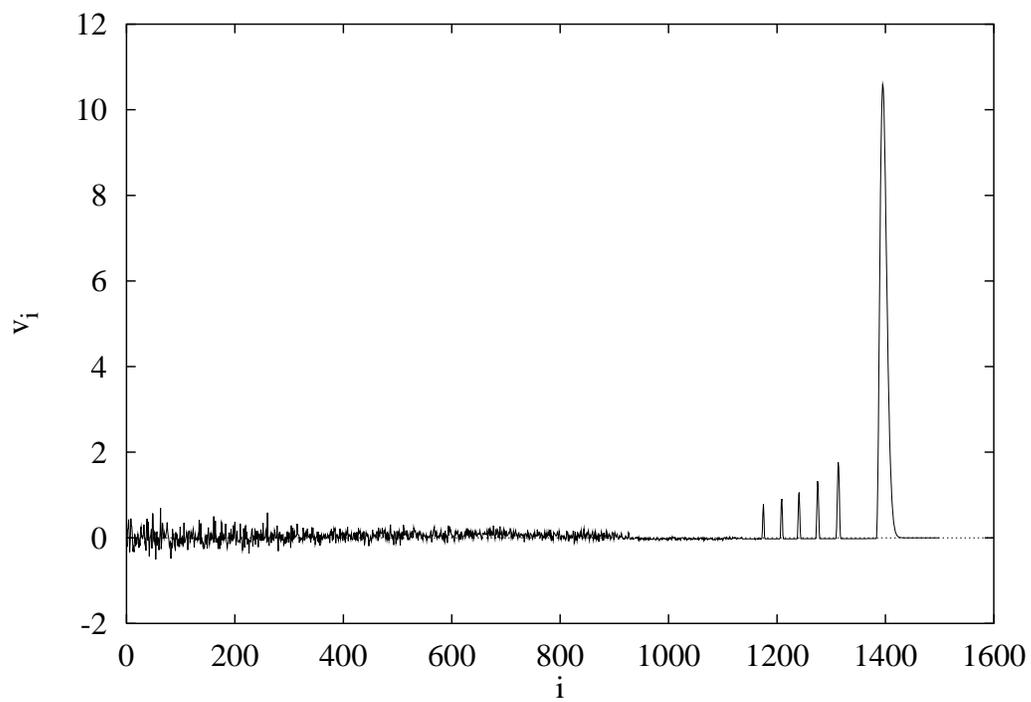,width=14cm}}
\caption[]{Velocities of the beads after $1.4\times10^{-2}s$.}
\label{fig2}
\end{figure}

In order to follow the evolution of the perturbation we have monitored
the evolution of the displacements $u_i$ and of the velocities $v_i$
of the beads versus the bead number $i$ as a function of time. A
snapshot of the solutions at a given time is shown in Fig.(\ref{fig1})
and Fig.(\ref{fig2}).  The bead displacements are characterized by a
front wave followed by inclined plateaus.  The plateaus consist of an
almost smooth curve with jumps.  It is worth to note how, in the plot
of the displacements, intervals with positive slopes typically
correspond to regions of detached particles while the negative slopes
correspond to particles in contact with each other.  The snapshot for
the velocities exhibits a structure of peaks with decreasing
amplitude. These peaks correspond to waves propagating with
monotonically decreasing velocities.  Each wave is, at its turn,
composed by several particles which are all in contact with each other
and moving in the same forward direction.  In between the peaks
particles are not in contact and propagate in the backward direction
 with decreasing velocities. In Fig.(\ref{fig2})
we also see the appearance of noise. This is a real numerical noise
which is reduced but never disappears when we decrease the time step
resolution. In fact, decreasing of one order of magnitude the
resolution allows us to observe another peak and so on and so
forth. In the ideal case of infinite resolution one would be able to
observe an infinite series of peaks behind the front. As we shall see
later this fact has no consequences on our conclusions since we will
discuss the region where no noise occurs and results can be
extrapolated to the noisy region.  Coming back to the analysis of
Fig.(\ref{fig2}) we can say that periodically the front looses energy
when its last particle is detached, i.e looses contact. This is the
only way in which the energy stored in the front wave can decrease.
In this way behind the front one has a chain of particles moving
backward with decreasing velocity.  These particles can eventually
contribute to the formation of peaks whose nature is different from
that of the front.  We will come back to this point later when we will
discuss the actual mechanism for the front formation.

It can be easily shown that the front wave propagates nearly 
at the sound velocity.  In order to do that we calculate analytically 
the speed of sound in the medium of the same density of our beads and 
we compare the results with the speed observed numerically for
several values of the elastic constant $k$: 
$10^8N.m^{-1}$, $10^7N.m^{-1}$ and $10^6N.m^{-1}$.

The calculation of the speed of sound is made as follows: if we
consider the beads as springs of spring constant $k$ and length $2R$
we get a dispersion relation that is:
\begin{equation}
\omega=2\sqrt{k \over m}\arrowvert\sin{qR}\arrowvert,
\end{equation}
where $q$ is the wave vector.
The speed of sound being defined by $c_s=\lim_{q \to 0}
{\omega \over q}$, we get
\begin{equation}
c_s=2R\sqrt{k \over m},
\end{equation}

Replacing all the parameters with their numerical values we have
$c_s=3170.47m.s^{-1}$ when $k=10^8N.m^{-1}$, $c_s=1002.59m.s^{-1}$
when $k=10^7N.m^{-1}$ and $c_s=317.05m.s^{-1}$ when
$k=10^6N.m^{-1}$. These speeds are systematically slightly greater
than the ones computed numerically which gave us $3160m.s^{-1}$,
$1000m.s^{-1}$ and $316m.s^{-1}$ respectively. These numerical values
have been obtained by counting the number of beads separating the
positions of the maxima of velocity in the front at two different
times. We cannot thus pretend an exact agreement with theoretical
values.

In order to have a better understanding of our results 
we compare our system with the harmonic chain consisting in a series of
springs with spring constant $k$. We integrated the
harmonic chain keeping the same initial conditions. 
After some algebra
we get the chain eigenfrequencies:
\begin{equation}
{\omega_l}^2=4{k \over m}\cos^2{l\pi \over 2N-1} \quad
1 \leq l \leq N-1
\end{equation}
We then obtain for the displacements:
\begin{equation}
\left\{
\begin{array}{l}
u_n(t)= \delta_0 + (-1)^n\sum_{l=1}^{N-1}C_l\sin{2nl\pi \over
  2N-1}\cos{\omega_l t} \quad 1 \leq n \leq N-1\\
u_0(t)= \delta_0
\end{array}
\right.
\end{equation}
where the $C_l$ are constants depending on the initial condition.

\begin{figure}[htbp]
\centerline{
        \psfig{figure=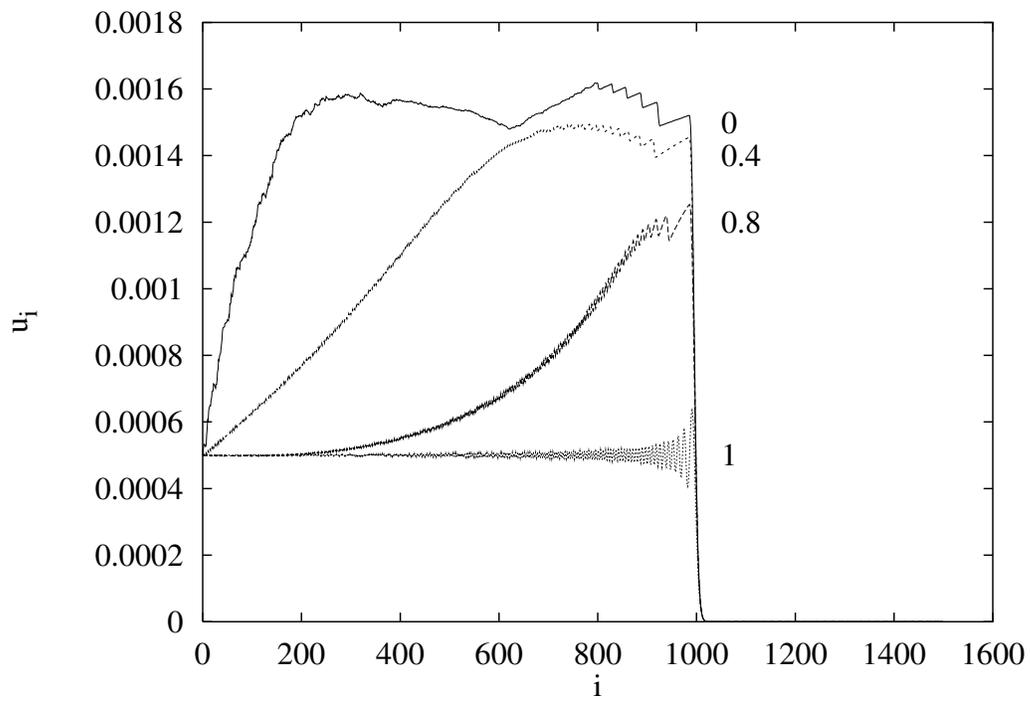,width=14cm}}
\caption[]{Curves of displacements for different values for the ratio
  of the elasticity constants. The ratio goes from one (harmonic chain) to
  zero (nonlinear chain). The propagation time is $10^{-2}s$. }
\label{fig3}
\end{figure}

This relation is shown in Fig.(\ref{fig3}) where we recognize the
well-known oscillations after the front due to the discretisation as
already discussed by Gibbs and being now what we call the ``Gibbs
phenomenon''.  We also plotted in Fig.(\ref{fig3}) what happens when
one tries to go continuously from the harmonic case to the no-tension
case by decreasing the spring constant of the springs when they are
under traction and keeping them equal to $k$ under compression. Each
curve corresponds to a different value of the ratio of the spring
constant under traction over $k$. We clearly observe a continuous
transition from the harmonic regime to the no-tension one as we
decrease the ratio of the spring constants.

\section{Wave formation}

In the previous section we have made some observations about the
displacement and velocity profiles for a fixed time of propagation.
Let us now consider the time evolution in order to understand 
how the peaks are created and how they propagate.

      \subsubsection{Scaling of the front shape}

\begin{figure}[htbp]
\centerline{
        \psfig{figure=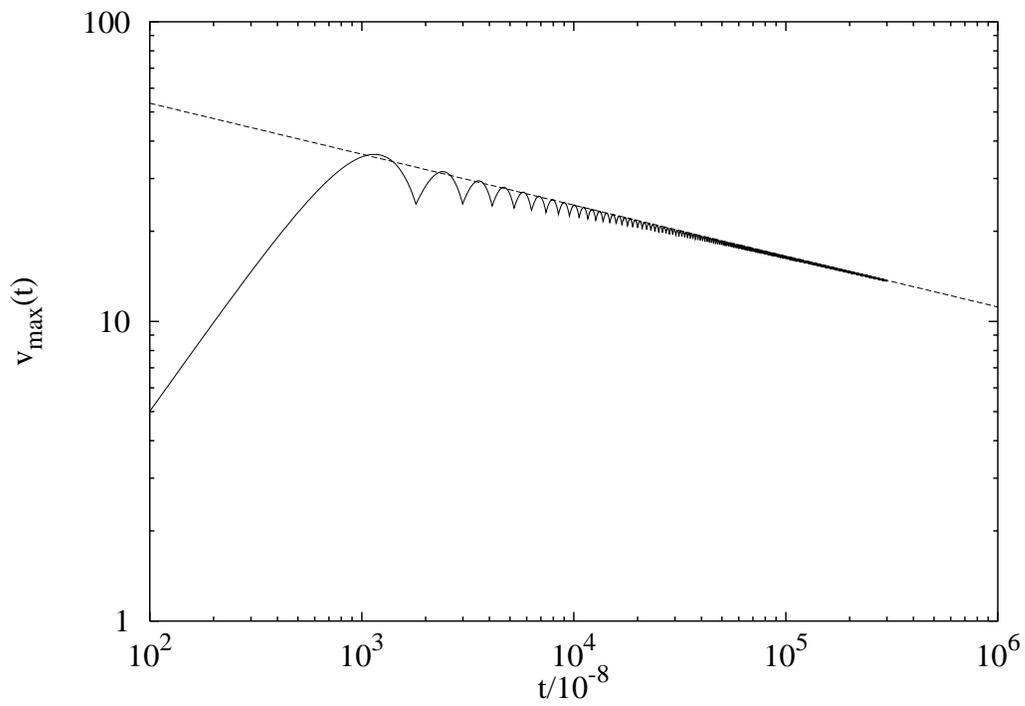,width=14cm}}
\caption[]{Double-logarithmic plot of the amplitude of the front wave versus
  time. After a fast increase it decreases and oscillates.  The dashed
  line corresponds to the function $f(x) \propto x^{-\alpha}$ which fits
  the envelope of the maxima of the amplitude curve.}
\label{fig4}
\end{figure}

We define the amplitude of the front wave as the maximum velocity of
the beads belonging to it. The curve representing this amplitude as a
function of time is shown on Fig.(\ref{fig4}).  One can see that the
amplitude decreases and oscillates. These oscillations are due to the
discretisation, i.e to the fact that the bead having the maximum
velocity keeps it during a finite time corresponding in
Fig.(\ref{fig4}) to the width of the oscillation. Initially the
fastest bead has a low velocity that increases towards a maximum with
time. Then this value decreases to a value lower than the initial
value. This behavior continues with the right neighbor of the
previous bead which then has the new maximum speed and keeps it during
a finite time with the same evolution as before. On a log-log scale we
find that the envelope of the maxima is well fitted by a power law
$A(t) \propto t^{-\alpha}$ with $\alpha \simeq 0.17$, consistent with
a $t^{-{1 \over 6}}$ behavior. This value of the exponent seems to be
universal since it is robust when changing the value of the elasticity
constant $k$ in the range $10^6 - 10^8 $ and $\delta_0$ in the range
$0.3 - 0.7 mm$.

\begin{figure}[htbp]
\centerline{
        \psfig{figure=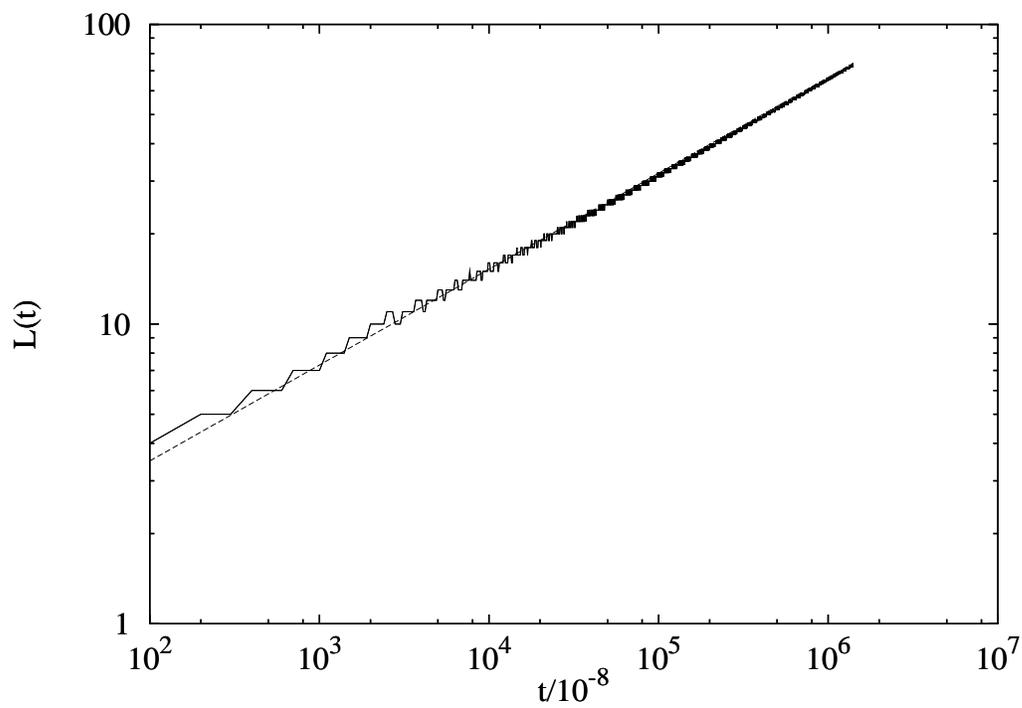,width=14cm}}
\caption[]{Double-logarithmic plot of the width of the front wave versus time. 
  The dashed line corresponds to the function $f(x) \propto x^{\beta}$
  showing an increase of the width with a power law.}
\label{fig5}
\end{figure}

We also define the width of the front wave by counting the number of
beads with velocities greater than zero belonging to the front.  For
this quantity we find a scaling law $L(t) \propto t^{\beta}$ with
$\beta\simeq 0.32$ (see Fig.(\ref{fig5})).  In this case the behavior
is consistent with $t^{1 \over 3}$ and is very robust with the same
universal character as for the scaling of $A(t)$.  The exponents
$\alpha$ and $\beta$ agree with the calculation made by J. Hinch
\cite{Hinch} using the conservation of kinetic energy of the front
wave ($\alpha^{\prime}={1 \over 6},\beta^{\prime}={1 \over 3}$).

\begin{figure}[htbp]
\centerline{
        \psfig{figure=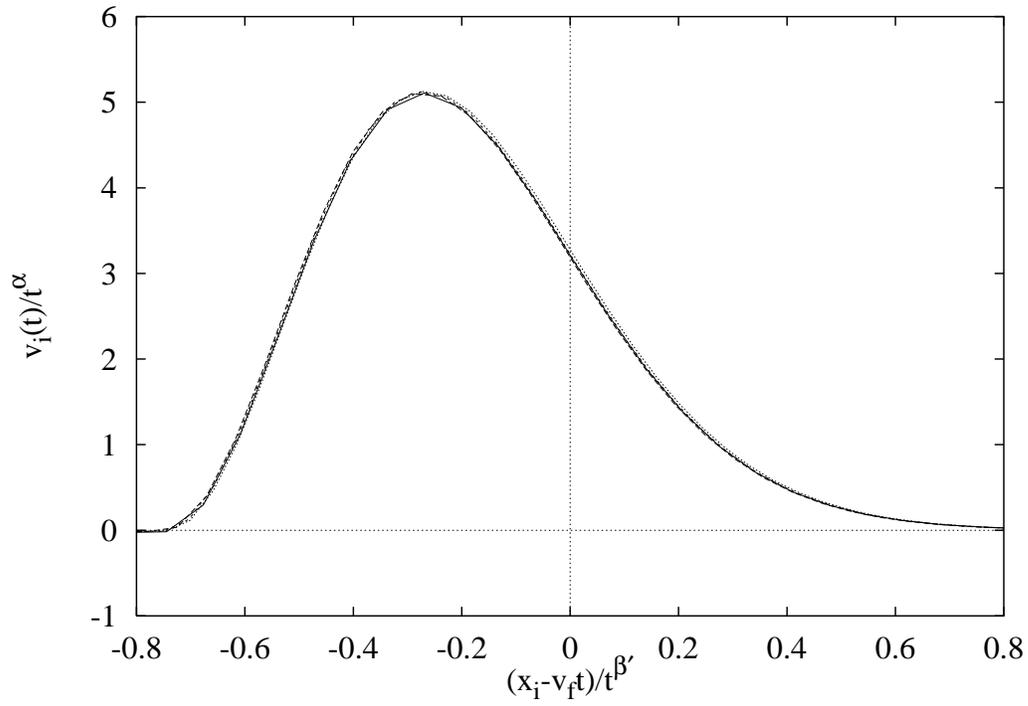,width=14cm}}
\caption[]{Collapse of the front wave shape to a single curve verifying
  the scaling law for the bead velocities in the front wave. We show
four curves corresponding to
$t=0.3\times10^{-2}s, 0.6\times10^{-2}s, 0.9\times10^{-2}s, 1.2\times10^{-2}s$.
$\alpha=0.17$, $\beta^{\prime}={1 \over 3}$ and $v_f=1001.5m.s^{-1}$.}
\label{fig6}
\end{figure}

The use of the two previous power laws enables us to find a scaling
relation for the front. By plotting $v_i(t) \over t^{-\alpha}$ versus
$x_i-v_ft \over t^{\beta^{\prime}}$ where $v_f$ is an adjustable
parameter with the dimension of a speed, we found that the curves
corresponding to different values of $t$ collapse on a single one as
shown on Fig.(\ref{fig6}). The collapse happens for a value of
$v_f=1001.5 m.s^{-1}$ which is very close to the sound speed
$1002.5m.s^{-1}$ ($k=10^7N.m^{-1}$). It is worth to note that $v_f$ is
the speed measured at the center of mass of the front and not at its
maximum. The best collapse is obtained for the values of $\alpha=0.17$
and $\beta^{\prime}={1 \over 3}$. The exponent $\beta$ changed to
$\beta^{\prime}$ in agreement with the theoretical results whereas
$\alpha$ is not exactly equal to $\alpha^{\prime}$ since its
calculation is based on kinetic energy conservation. In the present
case energy is dissipated from the front wave by the effect of
detachment of particles which induce a correction to
$\alpha^{\prime}$.
 We can therefore write the following scaling-law for the elastic front:
\begin{equation}
 {v_i(t) \over t^{-\alpha}}=f\Big({x_i-v_ft \over t^{\beta^{\prime}}}\Big)
\label{scaling}
\end{equation}
where $f$ is a scaling function that has also been calculated \cite{Hinch} and $v_f=1001.5m.s^{-1}$.

This scaling relation allows us to describe the self-similarity of the
front wave shape. Another interesting feature coming from the scaling
relation is that the velocity of the front is defined at the center of
mass of the beads belonging to it. The other points on the front shape
move with different velocities. Each point behind the center of mass
moves with a velocity smaller than the sound speed whereas the points
on the part preceding the center move with a larger velocity. Thus the
front wave is divided into a subsonic and a supersonic part. This kind
of wave is very different from the solitonic ones discovered by
Nesterenko \cite{Nesterenko} 
where it was found that the wave travels with a constant shape and a
constant speed for all its points.

   \subsubsection{Emergence of the secondary peaks}

Let us now focus our attention on the secondary peaks propagating
behind the elastic front.  The particles detached from the front move
backwards and their energy does the same (the system is globally
conservative).  This energy propagating backwards is then transferred
from particle to particle until it reaches the left boundary where it
is reflected and starts to propagate forward. This is the beginning of
the propagation of a peak.  The process then continues by means of
detachments of particles from this peak contributing in this way to
the formation of new peaks.
 
\begin{figure}[htbp]
\centerline{
        \psfig{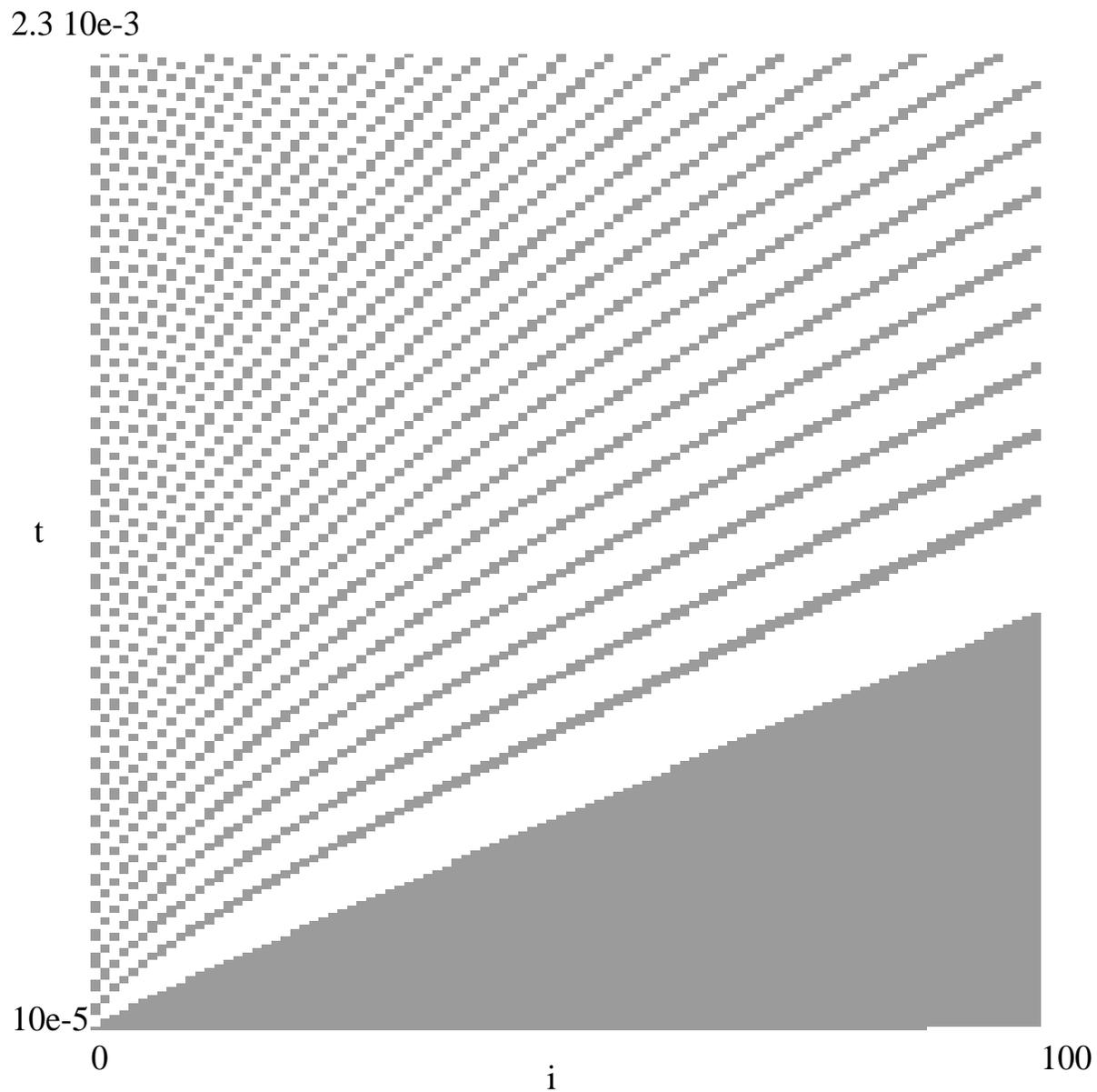}}
\caption[]{Spatio-temporal evolution of beads contacts. The first one
  hundred contact points of the chain are on the horizontal axis and
  time is on the vertical axis. The distribution of contacts is
  computed every $10^{-5}s$. Here we started at $10^{-5}s$ till
  $2.5\times10^{-3}s$. A white square means that there is no contact between
  two beads whereas a grey one means that the beads are in contact.}
\label{fig7}
\end{figure}

In order to analyze in detail the evolution of the peaks
we looked at the spatio-temporal structure
of the contacts of the first one hundred beads of the chain. 
This can be seen in Fig.(\ref{fig7}): 
a grey dot means that there is a contact between
two beads whereas a white dot means that they are not in contact. The
computation of the contact distribution of the beads is done every
$10^{-5}s$ up to a propagation time of $2.5\times10^{-3}s$. 
At $10^{-5}s$ almost all the beads are under compression.  Then this
compression chain moves to the right letting empty spaces between the
first beads of the chain. This corresponds to the formation of the
front wave. As the compression chain travels (the large grey triangle
on the right lower corner of the plot) one can observe grey dots
appearing at the beginning of the chain. This means that a new
compression wave is being created. This wave is much thinner than the
front wave and is represented on the plot by an almost straight
curve with a slope slightly above the slope of the front showing that
this first peak travels slower than the front. It can be seen that the
other peaks are also formed at the begining of the chain all making an
almost straight grey line with a slowly increasing slope which means
a decreasing speed. It is easy to understand that the existence of these
peaks is due to reflections on the first bead which is maintained
fixed playing the role of a reflecting wall. Hence without noise there
should be more than five  peaks in fig.(\ref{fig2}): 
all the peaks behind these five ones have been destroyed by a numerical
noise and one should see a succession of peaks with decreasing
amplitude until the beginning of the chain. One more important point
emerging from the spatio-temporal pattern is that beads which do not
belong to a peak are not under compression and are thus in a ballistic
regime as one can see from the white spaces separating the peaks in
Fig.(\ref{fig7}).

\section{Conclusion}

We have studied the response of a chain of beads to a small
displacement when no static force is applied to its ends. We
observe the propagation of waves very different from the solitons
which can be seen in a hertzian chain. 

The non-linearity of the problem lies in the contact law used: a 
step function which indicates that there is
no force when two beads are not in contact.

By keeping the left end of the chain fixed we observed an interesting
bouncing phenomenon. In addition to the elastic front, secondary peaks
of decreasing amplitude are generated at the left end of the
chain. These peaks correspond to several beads in compression
traveling in the forward direction whereas the beads in between the
peaks travel in the backward direction without touching each other.
We have found an interesting scaling law for the velocity of the beads
belonging to the elastic front.  This law which has also been derived
analytically by J.Hinch \cite{Hinch} seems to be universal since it
does not depend on the parameters of the problem.

Starting form these results it would be interesting to look at a chain
submitted to other types of perturbations. Besides, it is
important to look at different networks (two or three dimensional) in
order to understand the effect of the network structure on wave
propagation.
   
\vspace{5mm}
\noindent{\bf \large Acknowledgments}

After completing the paper we became aware that J. Hinch \cite{Hinch} 
studied a very similar problem getting very similar results. 
We thanks him for having shown to us his analytical results. We are
also grateful to S. Roux for many interesting discussions.
V.L. has been supported by Contract No. CEE ERBFMBICT 961220.

%\begin{figure}[htbp]
%\centerline{
%        \psfig{figure=viz.eps,width=14cm}}
%\caption[]{Enlargement of the previous figure showing the five
%  secondary peaks behind the front wave. One can see that the beads
%  between the peaks move with negative velocities.}
%\label{fig3}
%\end{figure}

%\begin{figure}[htbp]
%\centerline{
%        \psfig{figure=enf.eps,width=14cm}}
%\caption[]{Energy decrease of the first wave versus time.}
%\label{fig7bis}
%\end{figure}

%\begin{figure}[htbp]
%\centerline{
%        \psfig{figure=ci.epsi,width=9cm,height=11cm}}
%\caption[]{Evolution of the ten first beads of the chain from $0s$ to
%  $9\times10^{-5}s$. White beads have velocities less or equal to
%  zero whereas black beads are moving with positive velocities. }
%\label{fig8}
%\end{figure}

\begin{thebibliography}{99}

\bibitem{JNB} H.M. Jaeger, S.R. Nagel and R.P. Behringer, {\em
  Rev. Mod. Phys.} {\bf 68}, 1259-1273 (1996).
\bibitem{Liu} C. Liu, {\em Phys. Rev. B} {\bf 50}, 782 (1994).
\bibitem{LiuNagel1} C. Liu and S.R. Nagel, {\em Phys. Rev. Lett.} {\bf
    68}, 2301 (1992).
\bibitem{LiuNagel2} C. Liu and S.R. Nagel, {\em Phys. Rev. B} {\bf 48}, 15646 (1993).
\bibitem{Sornette} S. Feng and D. Sornette, {\em Phys. Lett. A} {\bf
    184}, 127 (1993).
\bibitem{Melin} S. Melin, {\em Phys. Rev. E} {\bf 49}, 2353 (1994).
\bibitem{Goddard} J. D. Goddard, {\em Proc. R. Soc. London} {\bf A
    430}, 105 (1990).
\bibitem{Nesterenko} V. F. Nesterenko, {\em J. Appl. Mech. and Tech. Phys.}
  {\bf 24}, 567 (1983).
\bibitem{LazaridiNesterenko} A. N. Lazaridi and V. F. Nesterenko, {\em
    J. Appl. Mech. and Tech. Phys.} {\bf 26}, 405 (1985).
\bibitem{CFF} C. Coste, E. Falcon and S. Fauve, {\em Phys. Rev. E}
  {\bf 56}, 6104 (1997).
\bibitem{Hinch} J. Hinch, {\em submitted to Proc. R. Soc. London}.
\bibitem{AA} M.K. Alves and B.K. Alves, {\em European Journal of
    Mechanics A/Solids} {\bf 16}, 103-16 (1997). 
\end{thebibliography}
\end{document}